\documentclass{article}
\usepackage[utf8]{inputenc}
\pdfoutput=1
\usepackage{graphicx}
\usepackage{verbatim}
\usepackage{amsmath}
\usepackage{appendix}
\usepackage{siunitx}
\usepackage{cite}
\usepackage{float}
\usepackage{hyperref}
    \hypersetup{
        colorlinks=true,
        linkcolor=blue,
        filecolor=blue,      
        urlcolor=blue,
        citecolor=blue
    }
\usepackage{array}
\usepackage[table]{xcolor}
\renewcommand{\arraystretch}{1.5}
\newcolumntype{s}{>{\columncolor[HTML]{AAACED}} p{3cm}}
\usepackage{booktabs}
\usepackage{multirow}
\usepackage{siunitx}
\usepackage{longtable}
\usepackage[left=4cm, right=4cm, top=2.5cm, bottom=2.5cm]{geometry}

\usepackage{authblk}
\title{Influence of individual factors on fundamental diagrams of pedestrians}
\author[1,*]{Sarah Paetzke} 
\author[1]{Maik Boltes} 
\author[1,2]{Armin Seyfried}
\affil[1]{Institute for Advanced Simulation, IAS-7: Civil Safety Research, Forschungszentrum Jülich, 52425 Jülich, Germany} 
\affil[2]{Faculty of Architecture and Civil Engineering, University of Wuppertal, 42285 Wuppertal, Germany}
\affil[*]{Corresponding author. E-mail address: s.paetzke@fz-juelich.de}
\begin{document}
\maketitle

\textbf{Highlights:}
\begin{itemize}
  \item Use of individual fundamental diagrams
    \item Analysis of the relationship between individual speed and headway
    \item Multiple linear regression to show how variables affect the individual speed
    \item Mixed models as an extension of regression analysis
\end{itemize}
\textbf{Abstract: }In recent years, numerous studies have been published dealing with the effect of individual characteristics of pedestrians on the fundamental diagram. These studies compared cumulative data on individuals in a group homogeneous in terms of one human factor such as age but heterogeneous in terms of other factors for instance gender. In order to examine the effect of all determined as well as undetermined human factors, individual fundamental diagrams are introduced and analyzed using multiple linear regression.
A single-file school experiment with students of different age, gender, and height is therefore considered. Single individuals appearing in different runs are analyzed to study the effect of human factors such as height, age and gender and all other unknown individual effects such as motivation or attention to the individual speed. The analysis shows that for students age and height are strongly correlated and, consequently, age can be ignored. Furthermore, the study shows that gender has a weak effect and other nonmeasurable individual characteristics have a stronger effect than height. 
In a further step, a mixed model is used as well as the multiple linear model.
Here, it is shown that the mixed model that considers all other unknown individual effects of each person as a random factor is preferable to the model where the individual speed only depends on the variables of headway, height, and all other unknown individual effects as fixed factors. 

\section{Introduction}
Fundamental diagrams describe the relationship between density, velocity or flow of people. They illustrate various traffic conditions including free flow, bounded traffic, maximum flow and congested traffic see e.g. \cite{Predtechenskii,Jelic2012,Holl:825757}. A differentiation is made between microscopic and macroscopic measurements. To represent the fundamental diagram the following variables are used. For the capacity of a system different combinations like flow and density $J(\rho)$ or flow and velocity $J(v)$ are applied. Furthermore, the velocity in dependence of the density $v(\rho)$ is used to relate travel times with the level of congestion.
Fundamental diagrams of various spatial structures, such as stairs, corridors or crossings, are different \cite{Weidmann1993, Fruin1971,Predtechenskii,BURGHARDT2013268,ZhangDiss,Vanumu}. Furthermore, human factors such as age, height, gender, culture, and motivation, external factors such as visibility or background music as well as different type of flows like uni-, bi-, or multidirectional streams all affect the fundamental diagram. 
These diagrams are examined in various studies, enabling comparisons of different cultures or people of different ages or gender. To date, factors such as height, gender, income, or culture have been studied at macroscopic and microscopic levels \cite{Ren2019a, CaoS2016, CaoS2018, CaoS2019, HuangS2018, Seyfried2010b, Zhang2014c, Subaih:873876,ZengG2018a,ZengG2018b, Subaih2019,Zhang2014b}. Further studies on single-file movement have considered external factors such as rhythm or background music \cite{Yanagisawa2012,ZengG2018b}, restricted visibility,  \cite{CaoS2019}, luggage and trolleys \cite{HuangS2018}, or properties of human locomotion such as step length and frequency \cite{WangJ2018,ZengG2018a, CaoS2018,YMa2018,Wang2018a,Song2013,Fujita2019}.

The various factors affecting the fundamental diagram illustrate the complexity of the quantitative description of pedestrian flow. The objective of this study is to introduce a new method to quantify the influence of individual characteristics of people on the fundamental diagram. For this purpose, regression analysis is used starting with simple linear regression and ending with mixed models. Moreover, we focus on how the methodology of the regression analysis changes the evaluation.  In order to explore the effect of different human factors, the study presented in this article is limited to the simplest system i.e., the movement of pedestrians in single file, as can be observed in queuing systems. In the following sections, we focus on single-file studies that consider age as a human factor.

Cao \cite{Cao2016} compared three groups, a younger, an older, and a mixed group. The group composed of younger people shows higher velocities than the mixed or older group for low and densities up to \mbox{$\rho=1.5$ [1/m]}. For high densities close to the stopping density, the speed of the mixed group is lower than that of the younger group. There is no comparative date for the group of older people. The fundamental diagram of the mixed group also has a more complex structure. This is illustrated by the fact that the diagrams cannot be transformed into each other by scaling the variables to dimensionless quantities. The diagram for the relationship between headway and velocity for the mixed group shows three regimes, the free, weakly constrained, and strongly constrained regime, whereas for the younger and older groups of people only two regimes can be observed. Between the younger and older group, on the other hand, only slight differences remain after the scaling procedure. 
A study by Ren et al. \cite{Ren2019a} confirms these results. The elderly group is slower than the group with young adults and also the comparison between the mixed group and the group composed of elderly people shows no differences. Moreover, in a headway vs. velocity diagram, a group of elderly Chinese and a group of French students \cite{Jelic2012} are compared and three regimes occur. Ren also concluded that the pedestrian dynamics are affected by factors such as age, heterogeneity of the group, and familiarity.
Zhang et al. \cite{ZhangJ2016} compared two groups of middle-aged individuals, with a low and good income. The fundamental diagrams for these groups are different but they show the same trend. One group with a good income and a higher number of female adults is more inactive and the participants prefer to maintain a greater distance from others or to keep pace with others. This group is more homogeneous, and more jams and stop-and-go waves occur. The other group with a low income and an approximately equal ratio of males and females is more active so the flow rate is higher. 

With regard to the age factor, Subaih et al. \cite{Subaih:873876} compared groups of different gender in experiments at high densities performed in Palestine and China. The authors concluded that older Chinese pedestrians walk as fast as young Palestinians but a group of younger Chinese people walk faster than younger Palestinians. 
A further study focusing on age is presented by Ziemer \cite{ZiemerDiss}. She analyzed experiments in schools by comparing students from the fifth grade with those from the 11th grade. The study shows that age has no effect on fundamental diagrams even if these groups have significantly different body heights. Groups that are heterogeneous in age also have no effect on the diagram. 
As to the question of whether the fundamental diagram has two or three regimes, we refer to Figures A18, A21e and A21g in \cite{ZiemerDiss}. These results indicate that the method of data binning is crucial.

From these six studies \cite{ZhangJ2016,Cao2016,Ren2019a,Subaih:873876,ZiemerDiss,Jelic2012}, it can be seen that age might have an effect on fundamental diagrams, but it does not normally. It depends on the age of the group whether it has an effect or not. Moreover, the discussion shows that beside the age, factors like culture, income, gender, and the homogeneity of the group composition could not be excluded. Also, there are indications that fundamental diagrams of heterogeneous groups have three regimes. But it depends on the binning method whether there are two or three regimes. 

Ren \cite{Ren2019a} shows that not only does age matters, but also the group composition in relation to the heterogeneity of the group in terms of gender or culture. Regarding Zhang \cite{ZhangJ2016}, the question arises whether the differences observed are affected by the different income or the gender composition of the group. In Subaih et al. \cite{Subaih:873876}, it is not clear whether age, culture, or the heterogeneous group composition with regard to gender have the main effect on the fundamental diagram. 

The discussion above gives a highly contradictory picture of how human factors affect the fundamental diagram of pedestrian dynamics. Even if these studies are performed under well-controlled laboratory conditions, the methodological problem is that even if a group is homogeneous in terms of one factor, it might be heterogeneous in terms of other factors. 

In the above-mentioned studies, measurements of the velocity and density of individuals are made. For the comparison of the fundamental diagrams, however, only the cumulative data of all individuals in the group are used. To solve this problem, individual fundamental diagrams are introduced and a multiple linear regression analysis is performed to study the effect of human factors. It is also taken into account that certain factors could be strongly correlated.

The structure of the paper is as follows. Section 2 describes the experimental setup, the measurement methods, and the data preparation. Then Section 3 deals with the regression analysis which includes the simple and the multiple linear regression and the mixed model. The last section highlights the conclusions and interprets the results. Finally further possible research steps are proposed.
\section{Materials and Methods}
\subsection{Experimental setup}
\begin{figure}[htp]
\centering 
\begin{minipage}[t]{0.45\linewidth} 
\centering
\includegraphics[width=5.9cm]{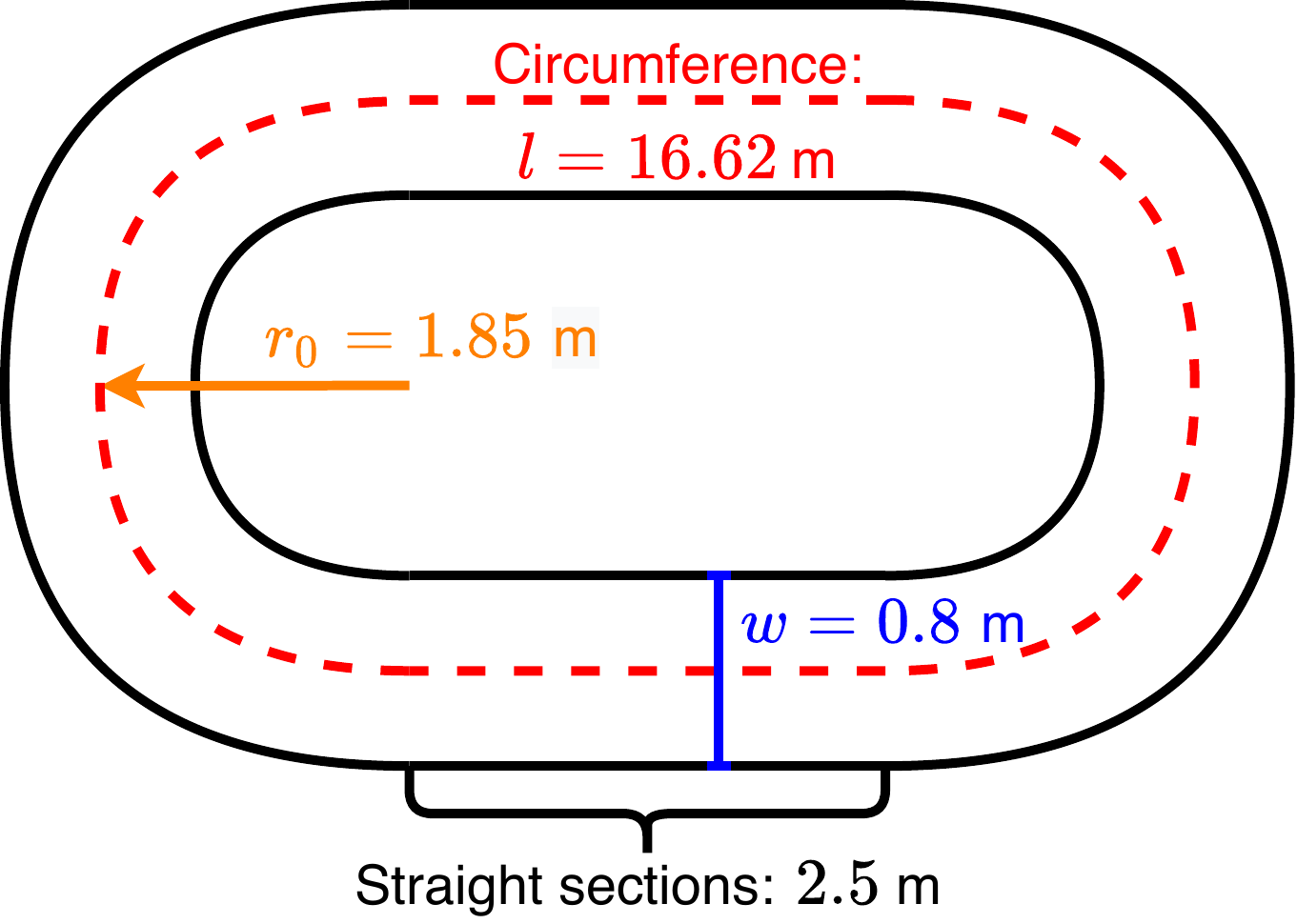}
\end{minipage}  
\hfill 
\begin{minipage}[t]{0.45\linewidth} 
\centering
\includegraphics[width=5.8cm]{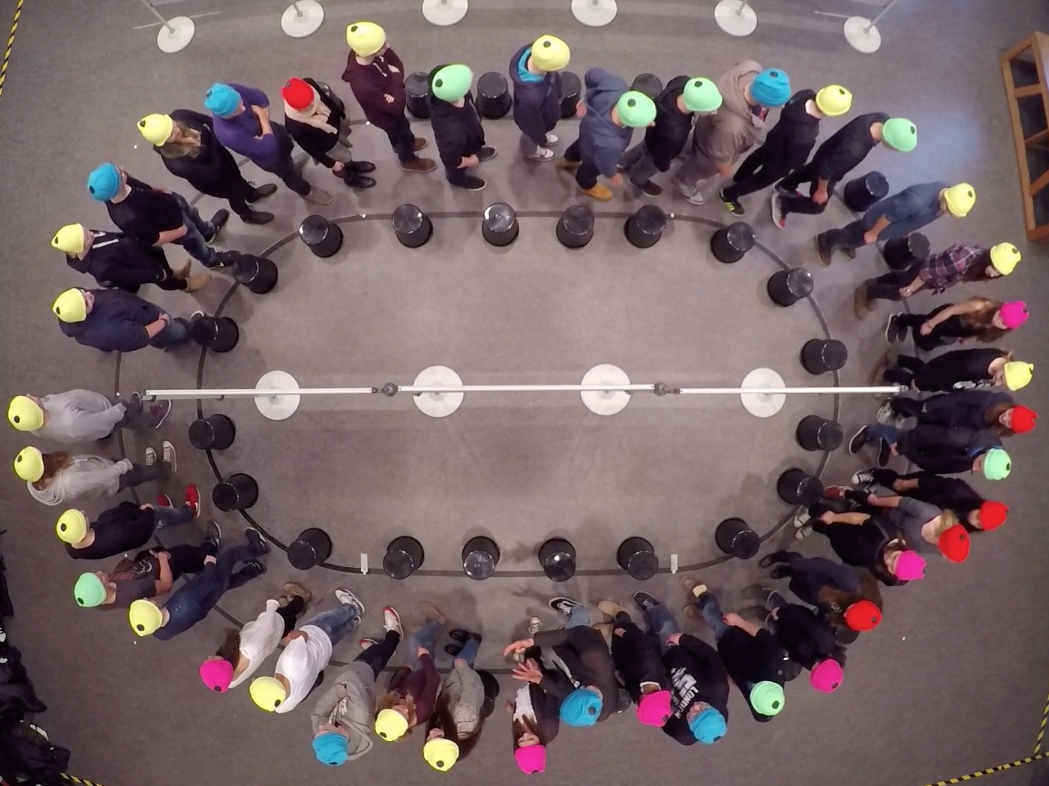}
\end{minipage}  
\caption{Single-file experiment at Gymnasium Bayreuther Straße (GBS) in Wuppertal, Germany. The figure on the left shows the oval path with the corresponding lengths. On the right is an overhead view of the experiment showing the students wearing colored caps on their heads. The colors indicate intervals of body heights.}
\label{fig:geometrysetup}
\end{figure}

In the present paper, a one-dimensional single-file experiment \cite{Data} performed at the school Gymnasium Bayreuther Straße (GBS) in Wuppertal, Germany in 2014 is analyzed. The spatial structure for the experiment is an oval path with a total length of the central line of \mbox{$l=16.62$ m}. The dimensions of the experiment can be seen in \autoref{fig:geometrysetup}. The oval path has a width of \mbox{$w=0.8$ m} and each straight section has a length of \mbox{$2.5$ m}. To reduce the complexity of the system the two-dimensional trajectories are mapped to one dimension. For this purpose, the participants' trajectories are projected on the middle line of the oval according to Ziemer \cite{ZiemerTraj}. Thus, only the change in movement direction over time is considered.

The participants are advised to walk behind each other without haste and without overtaking. In total, $118$ different students participated in the experiment, with around \mbox{$46 \ \%$} male pedestrians. Each subject is given a main ID in order to identify a particular student in different runs. This identification and the assignment of gender is realized manually. The subjects are from fifth and 11th grade. The younger students are aged between 11 and 12. The older ones are about 17 or 18. The students wear colored caps on their heads, which are used on the one hand to extract the trajectories and on the other hand to indicate different intervals of body height \cite{Boltes:62385, PeTrack}. In total, there are five different intervals to analyze the effect of height. The average height of younger students is \mbox{1.48 m $\pm$ 0.04 m} and that of older students is \mbox{1.76 m $\pm$ 0.07 m}.  The students are introduced to the topic in teaching units. See \cite{ZiemerDiss} for further information about the experiment.

A total of $31$ runs are performed with different global densities $\rho_{gl}\in[0.32,3.20]$[1/m] which are calculated by \begin{equation}\label{eq:globaldensity}
\rho_{gl}=N/l_m \ ,
\end{equation} 
the number of persons $N$ in one run in the measurement area and the length of the measuring area $l_m=15.62$ m because unrolled one-dimensional data are considered. An overview of the global densities for individual runs can be seen in \autoref{table:1}. 
\begin{table}[!ht]
\scriptsize
\setlength{\tabcolsep}{5pt}
\renewcommand{\arraystretch}{1.7} 
\begin{tabular}{cccccccccc}
\toprule
     &  {Run} & {Duration} & {I$_{min}$} & {I$_{max}$} & {$N$} & {$m$} & {$f$} & {$\rho_{gl}$} & {$\overline{\delta_t}\pm \sigma$}\\
    \midrule
  GBS-5th grade  & 1101  & 93.75 & 7 & 77.60 & 16 & 8 & 8 & 1.02 & 2.01 $\pm$ 0.38 \\
                                  & 1102  & 123.12 & 0 & 123.12 & 50 & 28 & 22 & 3.20 & 1.06 $\pm$ 0.82  \\  
                                  & 1103  & 167 & 2 & 167	& 40 &	20	 & 20 &	2.56  & 2.02 $\pm$ 0.73 \\
                                  & 1104  & 138.52 & 0 & 138.52 	& 32 &	20	 & 12 &	2.05  & 2.50 $\pm$ 0.51 \\
                                  & 1105  & 156.16 & 12 & 105 	& 10 &	6	 & 4 &	0.64  & 1.72 $\pm$ 0.49\\ 
                                  & 1106  & 119.08 & 0.4 & 119.08	& 24 &	13	 & 11 &	1.54  & 2.48 $\pm$ 0.46\\ 
                                  & 2101  & 77.64 & 6 & 77.64 & 24 &	13	 & 11 &	1.54 & 1.99 $\pm$ 0.39 \\ 
                                  & 2102  & 123.12 & 6 & 123.12	& 5 &	0	 & 5 &	0.32 & 2.00 $\pm$ 0.54\\ 
                                  & 2103  & 137.28 & 10 & 137.28	& 11 &	4	 & 7 &	0.70 & 1.99 $\pm$ 0.73\\ 
                                  & 2104  & 155.40  &  0 &  155.40	& 34 &	16	 & 18 &	2.18  & 2.72 $\pm$ 0.69\\
                                  & 2105  & 118.64  & 12 & 118.64	& 16 &	8	 & 8 &	1.02  & 2.04 $\pm$ 0.37\\
                                  & 2106  & 75.16 & 0.8 & 75.16	& 28 &	18	 & 10 &	1.79  & 1.10 $\pm$ 0.45\\ 
                                  & 2107  & 44.60 & 4 & 44.60	& 27 &	18	 & 9 &	1.73  & 2.44 $\pm$ 0.71\\ 
                                  & 2108  & 89.36 & 8 & 89.36	& 14 &	5	 & 9 &	0.90  & 1.09 $\pm$ 0.23 \\ 
                                  \midrule
     GBS-11th grade  & 1201 &  134.80  &  0 & 100 & 39 & 15 & 24 & 2.50 & 1.27 $\pm$ 1.06 \\
                                  & 1202 & 78.76  & 6 & 69.04 & 5 & 1 & 4 & 0.32  & 0.99 $\pm$ 0.39 \\
                                  & 1203  & 138 & 0 & 108 & 33 &	14	 & 19 &	2.11 & 1.10 $\pm$ 0.61 \\ 
                                  & 1204  & 107.72  & 12 & 100 	& 12 &	5	 & 7 &	0.77  & 1.48 $\pm$ 0.41\\
                                  & 1205  & 114.36 & 2 & 96 	& 23 &	9	 & 14 &	1.47  & 1.14 $\pm$ 0.43\\ 
                                  & 1206  & 91.60 &  12 & 81.20 	& 15 &	5	 & 10 &	0.96  & 1.83 $\pm$ 0.26\\ 
                                  & 1207  & 97.84 & 4 & 76 	& 21 &	5	 & 16 &	1.34  & 1.55 $\pm$ 0.40\\ 
                                  & 2201  & 103 & 8 & 99.04	& 5 &	3	 & 2 &	0.32  & 1.39 $\pm$ 0.32\\ 
                                  & 2202  & 109.40  & 0 & 72 	& 39 &	17 & 22 &	2.50  & 1.31 $\pm$ 0.63\\
                                  & 2203  & 118.20 & 8.80 & 111.44 	& 9 &	3	 & 6 &	0.58  & 1.31 $\pm$ 0.18 \\ 
                                  & 2204  & 121.76  & 0 & 96 	& 29 &	13	 & 16 &	1.86  & 1.17 $\pm$ 0.41 \\
                                  & 2205  & 133.88 & 6 & 122.32	& 15 &	6	 & 9 &	0.96 & 1.75 $\pm$ 0.27\\
                                  \midrule
 GBS-5th + 11th grade & 1301  & 142.84  & 0 & 104 & 42 & 20 & 22 & 2.69  & 0.81 $\pm$ 0.17\\
                                  & 1302 &  140.92  &  0 & 104 & 44 & 25 & 19 & 2.82 & 1.02 $\pm$ 0.73 \\ 
                                  & 1303  & 91.08   & 4 & 89.44	& 5 &	2	 & 3 &	0.32 & 1.13 $\pm$ 0.13 \\
                                  & 1304  & 134.12  & 0 & 104 & 33 &	17	 & 16 &	2.11  & 2.12 $\pm$ 1.12\\
                                  & 1305  & 76.96 & 4 & 70.68 & 11 &	5	 & 6 &	0.70 & 2.36 $\pm$ 0.67\\ 
                                  \bottomrule
  \end{tabular}
  \centering
  \caption{The columns from left to right show a detailed overview of the runs for the different groups of students from fifth grade, 11th grade and both fifth and 11th grade and their general properties. Column by column from left to right, the following information is included: The number of the run, the duration of a run in seconds, the interval of the average velocity in seconds, the total number of person in a run, the number of male and female pedestrians in a run, the global density and the mean values for individual specific time-steps and their standard deviation in seconds.}
\label{table:1}
\end{table}

\subsection{Measurement methods}
Based on the one-dimensional trajectories gained by tracking the head from video recording, the individual velocity and density as well as the headway are calculated using the software JPSreport \cite{JuPedSim}. For the one-dimensional case, the position of one individual $i$ at time $t$ is defined by $x_i(t)$ whereby $t$ is in the time interval $[\mbox{0, Duration}]$.

The headway $h_i(t)$ of student $i$ at time $t$ is calculated by \begin{equation}\label{eq:headway}
h_i(t)=x_{i+1}(t)-x_{i}(t) \ ,
\end{equation} which describes the distance between the centers of the heads $x_i(t)$ and $x_{i+1}(t)$ whereby $x_i(t)$ is the coordinate of pedestrian $i$ at time $t$ and $x_{i+1}(t)$ is the position of the person $i+1$ at time $t$ walking in front of person $i$.

\begin{figure}[htp]
    \centering
    \includegraphics[width=5cm]{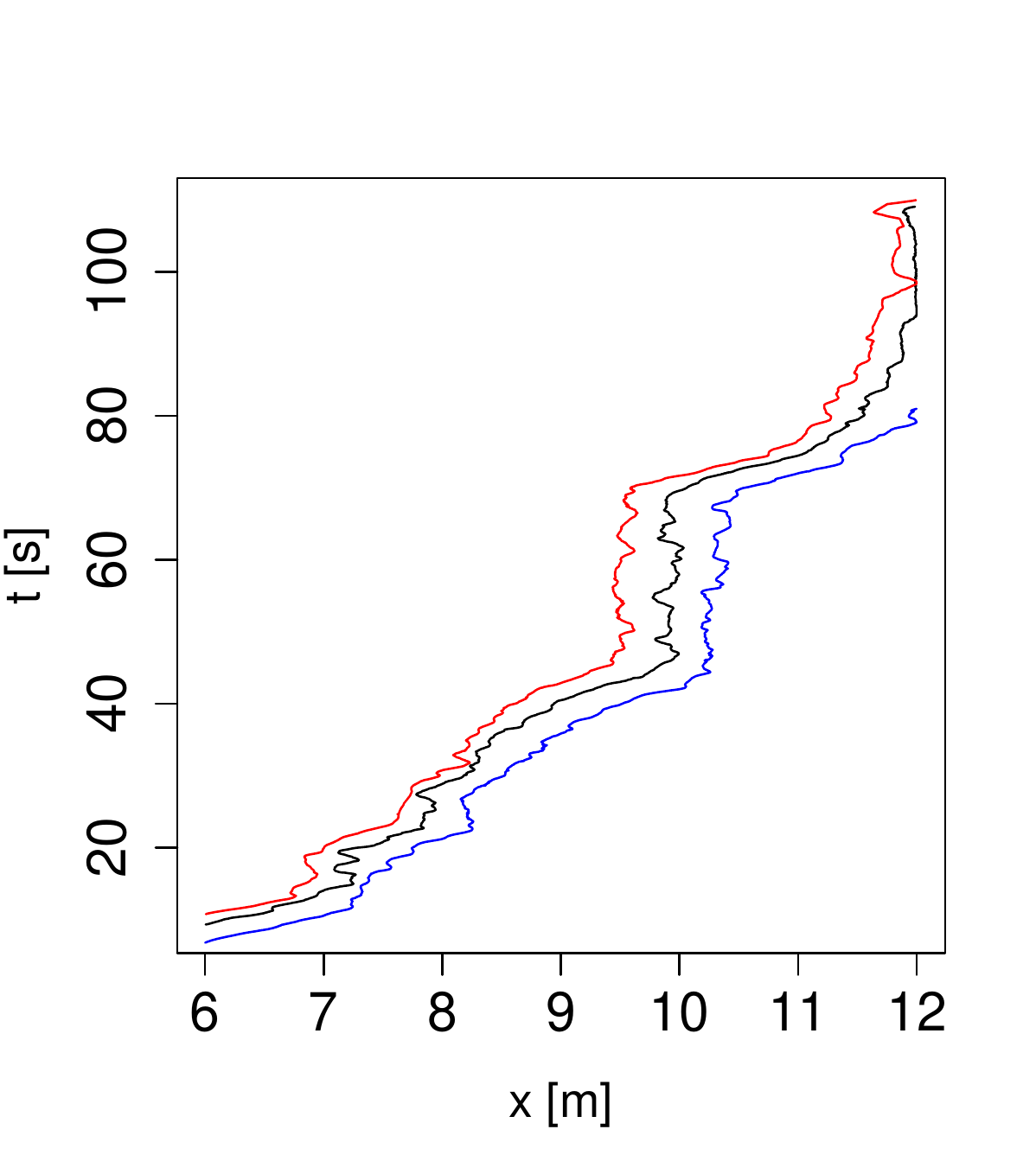}
     \caption{Walking position over time for three different individuals that show movements in positive and negative $x$-direction.}
   \label{fig:trajvelocity}
\end{figure}

The individual speed is also calculated using the software JPSreport. It applies: 
\begin{equation}\label{eq:indvelo}
v_i(t)=\frac{x_i(t+\frac{\Delta t}{2})-x_i(t-\frac{\Delta t}{2})}{\Delta t} \ .
\end{equation}
It should be noted that the default value $\Delta t$ which describes the difference between two time points is selected in such a way that the oscillations of the original trajectories caused by the movement in steps resulting in one dimension in a microscopic periodic speed pattern are smoothed out and so these do not have to be taken into consideration when analyzing the autocorrelation of the speed. Thus considering \cite{Tordeux2016WhiteAR}, \mbox{$\Delta t=0.8$ s} has been selected. The fixed value is a good assumption. The intended direction of the students is also included.
The reason for this is that, as can be seen in \autoref{fig:trajvelocity}, significant oscillations occur at a low speed.
Even if a pedestrian stops, their head moves, for example as a result of changing the leg they are standing on and thus the trajectory shows movement also in a negative $x$-direction. 
Hence, negative velocities can also be observed in the runs. This is shown, among other things, in the interval of \mbox{$x \in [9.5,10.5]$[m]} in \autoref{fig:trajvelocity}.

\subsection{Data processing}
\begin{figure}[htp]
    \centering
\includegraphics[width=9cm]{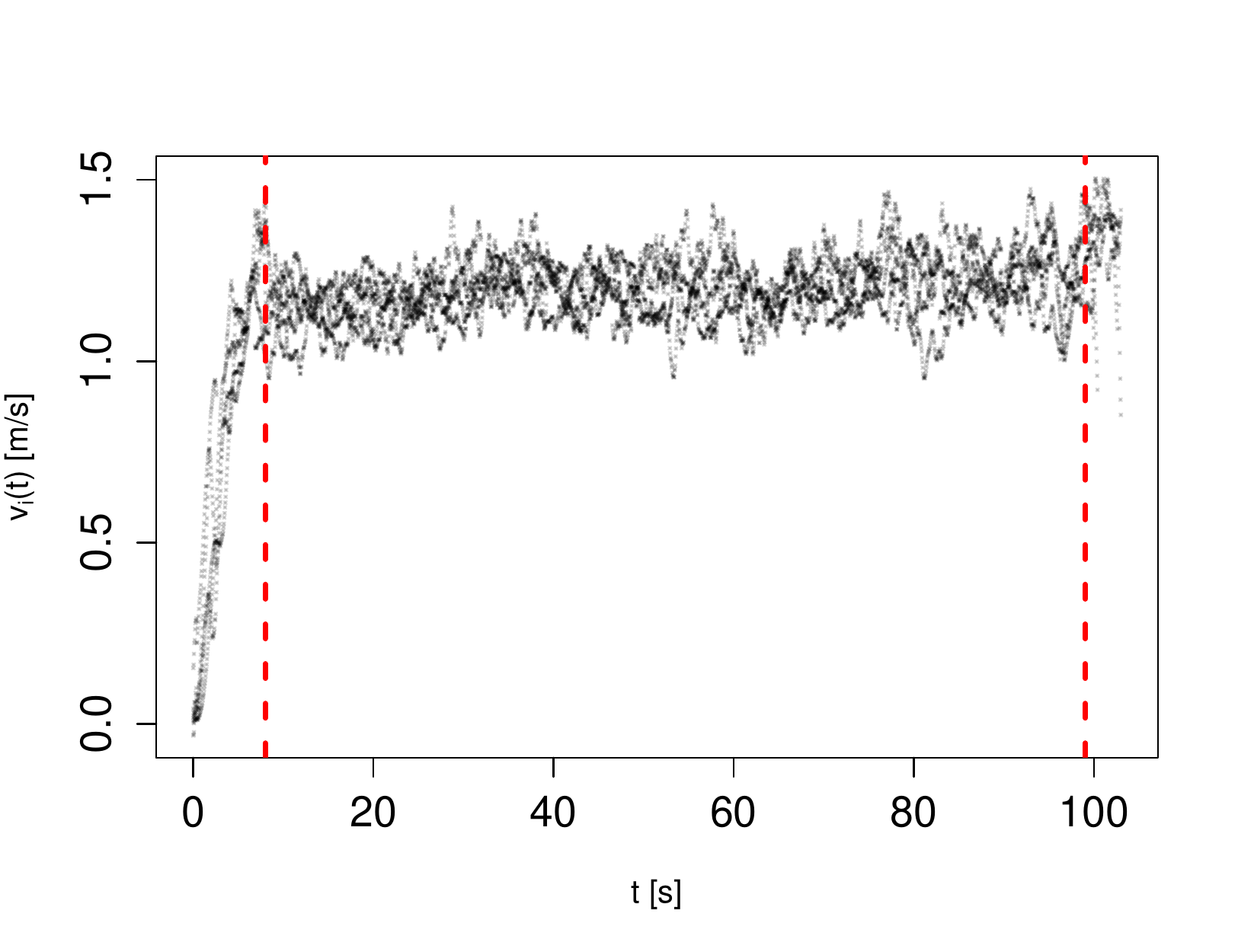}
\caption{Illustration of the steady state in run 2201, whereby the vertical red lines represent I$_{min}$ and I$_{max}$.}
\label{fig:steadystate}
\end{figure}

During the analysis, only the manually selected steady state which is chosen by looking at the average speed of the given run is considered. I$_{min}$ describes the time at which the average speed is reached for the first time and I$_{max}$ the time at which the average speed is reached for the last time (see \autoref{table:1}).   
In \autoref{fig:steadystate}, the straight red lines illustrate the boundaries of the steady state.

To ensure independent measurements in the regression analysis, the data are reduced by taking into consideration the autocorrelation. Here, we examine how strongly the observations of the individual speed depend on each other in the case of a time lag. To guarantee that the data are statistically independent for each individual, the speed values only are considered when the autocorrelation function applies \mbox{$r_{i,\tau} < 0.3$} for the first time. This decision is applied to the length of the individual time steps between the observation points in each run, for each individual.
Thus, the analysis is based on individual-specific $\delta_t$ time steps. \autoref{table:1} shows the mean values $\overline{\delta_t}$ of the different students in each run and their standard deviation. 

Now the data set can be used for further data analysis, including the regression analysis and the analysis of the individual fundamental diagrams but in order to obtain a higher number of data for each student for the study, the data from one individual in experimental runs of different densities are combined. Due to the link between the different runs and the individual-specific $\delta_t$ time steps, the number of observations $l=1,...,n_i$, for each individual $i$, differs.
Using the main ID of each person, we documented which runs with different densities an individual is involved in. In addition, the data for each individual is based not only on the different densities but also on various individual velocities and neighboring pedestrians as well as others preceding or following them. Thus, our observations are not only the result of a simple run and its composition but probably also represent a characteristic individual property such as preferences for certain individual velocities or for certain distances based on different neighboring students.

\section{Results}
\label{RegressionAnalysis}
\subsection{Structure of individual fundamental diagrams}
\label{Data selection}
In this section, the relationship between speed and headway is analyzed. In addition, we examine which factors significantly affect the speed and which can possibly be ignored.

\begin{figure}[h]
    \centering
\includegraphics[width=9cm]{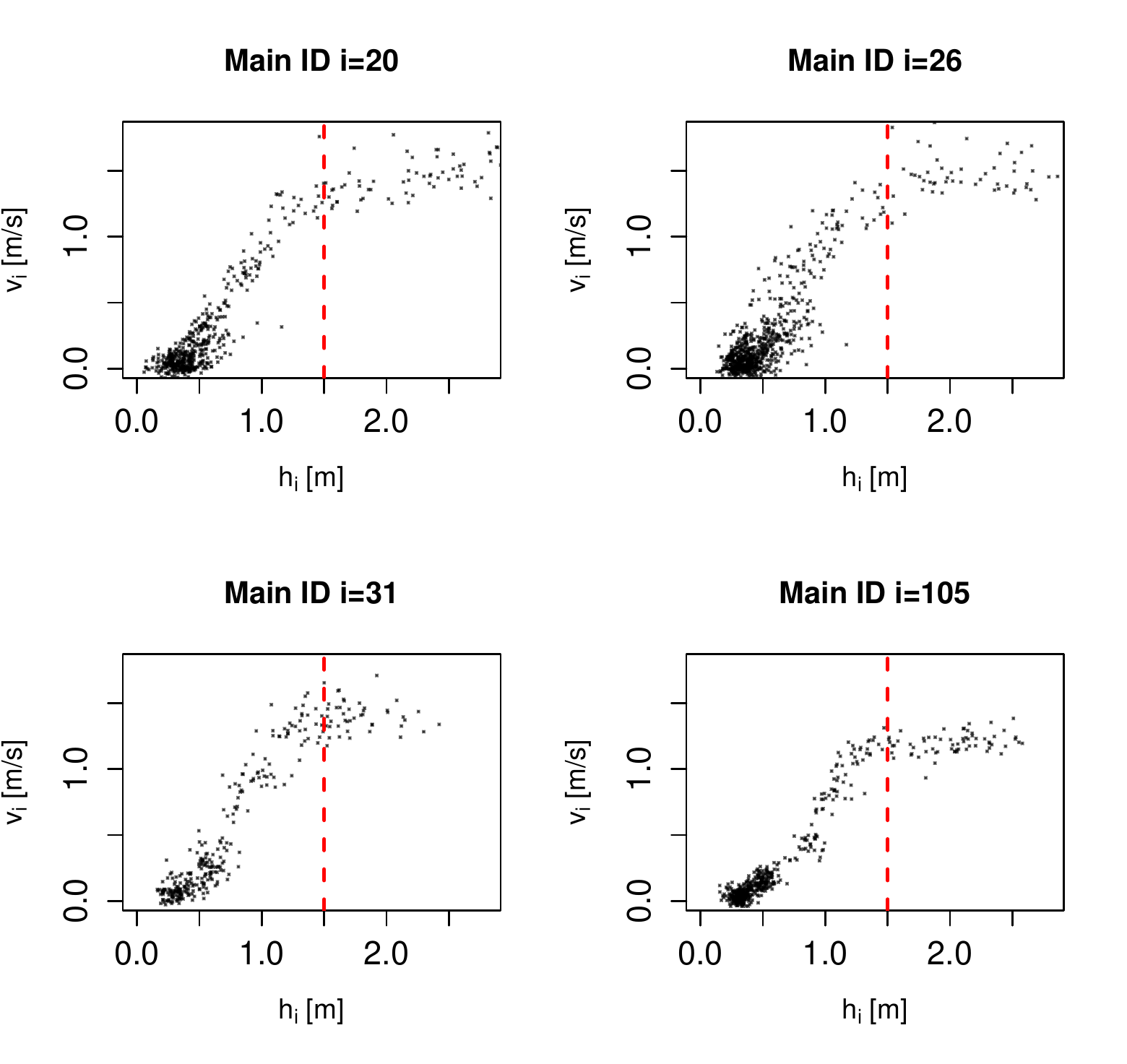}
     \caption{Headway vs. individual speed diagrams for four different main IDs to illustrate various linear sections and that the free flow area starts at \mbox{$h_i \approx 1.5$ m}.}
   \label{fig:HeadwayVeloFull}
\end{figure}

The diagrams in \autoref{fig:HeadwayVeloFull} illustrate exemplary the relationship between headway and the individual speed for a certain main ID of one individual. The data show that there are different regimes. However, it is not clear whether these are two or three different regimes but, in general, the free-flow branch starts clearly at a headway of \mbox{$h \approx 1.5$ m}. The beginning of the area selected for the free speed is supported by studies by Ziemer \cite{ZiemerDiss} and Cao \cite{CaoS2016}, who also examined younger age groups. 

We decided to study the branch where the speed is affected by neighboring pedestrians. Accordingly, the effect on the free speed is not analyzed. 

\subsection{Linear regression analysis and analysis of scattering}
\label{LinearRegression}
This section is about the research question whether individual speed-headway functions show that gender and age have an influence. A linear model is used as the simplest scenario for this study and the section for which
the analysis is performed to examine the effect of independent variables on the individual speed for each individual $i$ is \mbox{$h_i < 1.5 $ m}.
First, a simple linear regression analysis is applied. Depending on the number of individual observations $n_i$, the following formula results:
\begin{equation}\label{eq:IndividualModelSimpleRegression}
    v_{l}=\beta_0+\beta_1 \cdot h_{l}+\epsilon_{l}, \mbox{ where } l=1,...,n_i \ .
\end{equation}
The speed is represented by $v_{l}$, and $h_l$ is the headway. In \autoref{eq:IndividualModelSimpleRegression}, $\epsilon_{l}$ describes the random experimental error which should has a small scattering and $\beta _0$ and $\beta_1$ are unknown regression coefficients. 

\begin{figure}[htp]
\centering
        \begin{minipage}[t]{0.48\textwidth}
        \centering
                \includegraphics[width=\textwidth]{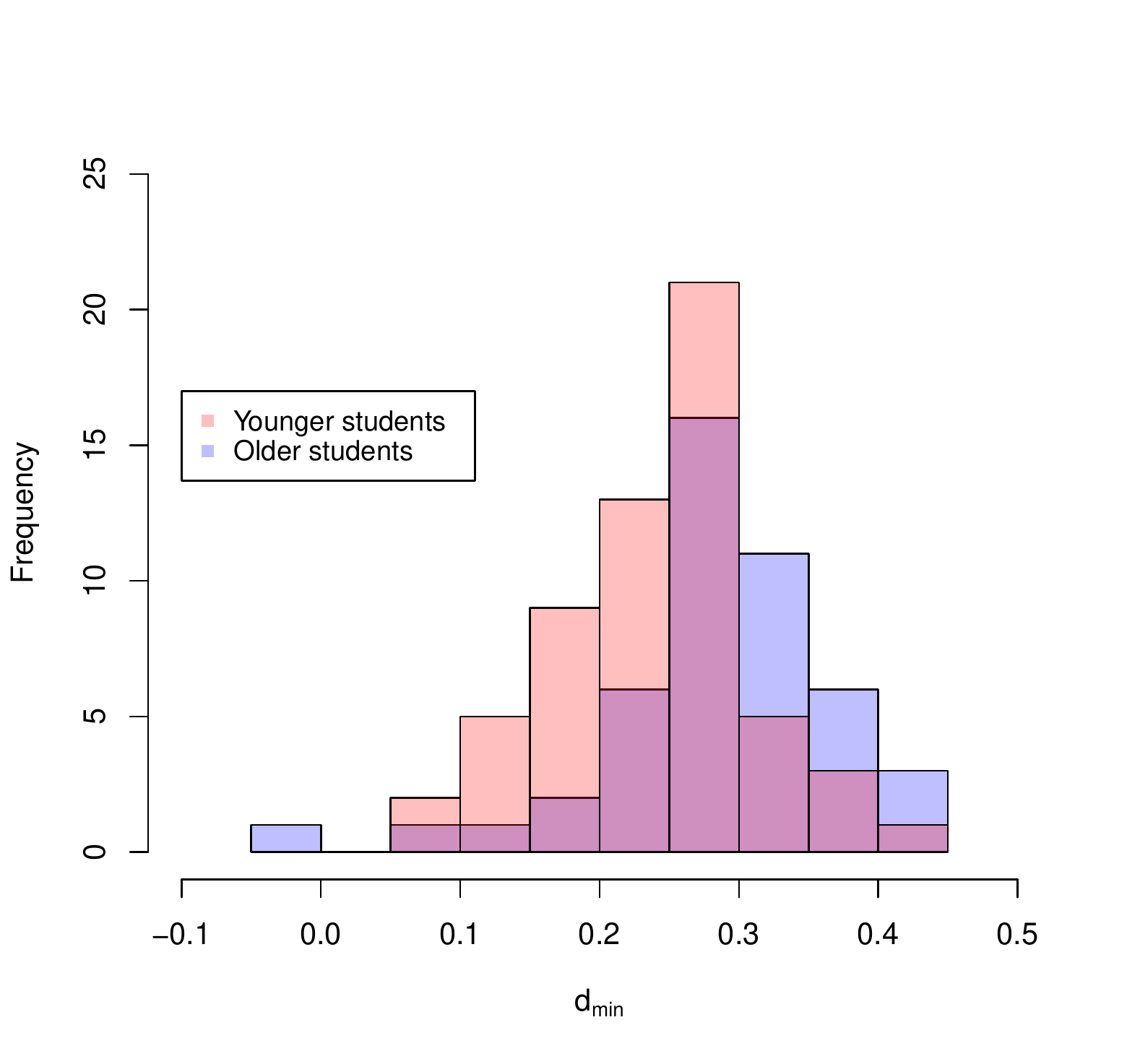}
        \end{minipage}\hfill
        \begin{minipage}[t]{0.48\textwidth}
        \centering
                \includegraphics[width=\textwidth]{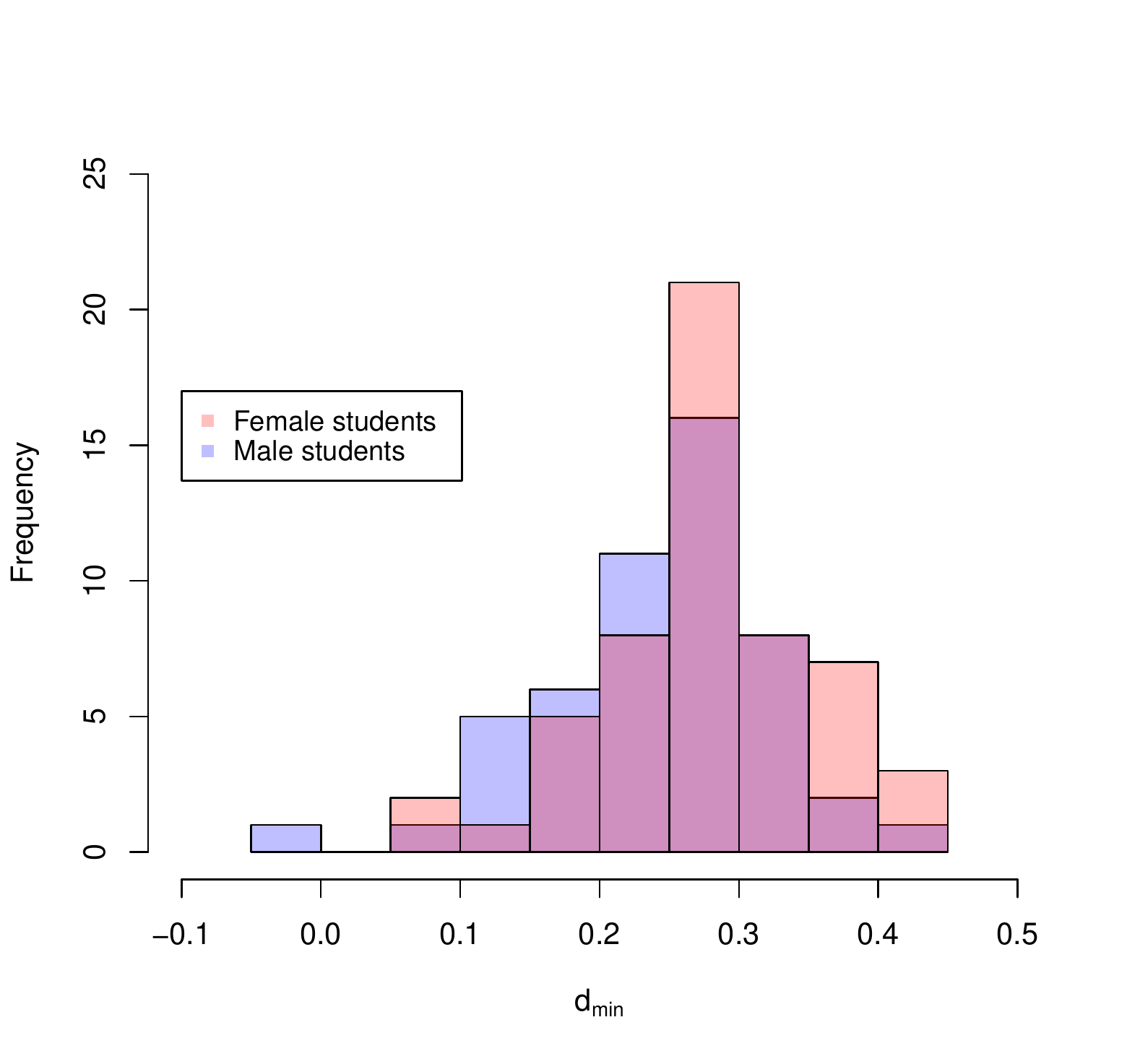}
        \end{minipage}
           \caption{The left-hand figure shows minimum distances for younger and older students and the right-hand one minimum distances for male and female students at $\hat{v}=0$.}
\label{fig:MinDistanceYoungOldMaleFemale}
\end{figure}

For a good fitting the values $\beta_0$ and $\beta_1$ need to be estimated.
\begin{equation}\label{eq:SimpleEstimated}
    \hat{v}=\hat{\beta}_0+\hat{\beta}_1 \cdot h \ ,
\end{equation}
whereby $\hat{\beta}_0$ and $\hat{\beta}_1$ are the estimated values that minimize fitting error.
Furthermore, \autoref{eq:SimpleEstimated} gives the slope of the regression line as $\hat{\beta}_1$ and by transforming the formula for $\hat{v}=0$, we obtain the minimum headway for each individual $i$:
\begin{equation}\label{eq:MinHeadway}
d_{min}=-\frac{\hat{\beta}_0}{\hat{\beta}_1} \ .
\end{equation}
\autoref{fig:MinDistanceYoungOldMaleFemale} illustrates that at $\hat{v}=0$, larger minimum distances occur for the older students. The mean values and standard errors are $\mu_{old} = 0.28\pm 0.01$ and $\mu_{young} = 0.24\pm 0.01$. The comparison between the male and female students shows that the minimum distance is slightly greater for female students $\mu_{male} = 0.25\pm 0.01$ and $\mu_{female} = 0.28\pm 0.01$. However, this difference is less pronounced. 

\begin{figure}[htp]
\centering 
        \begin{minipage}[t]{0.48\textwidth}
        \centering
                \includegraphics[width=\textwidth]{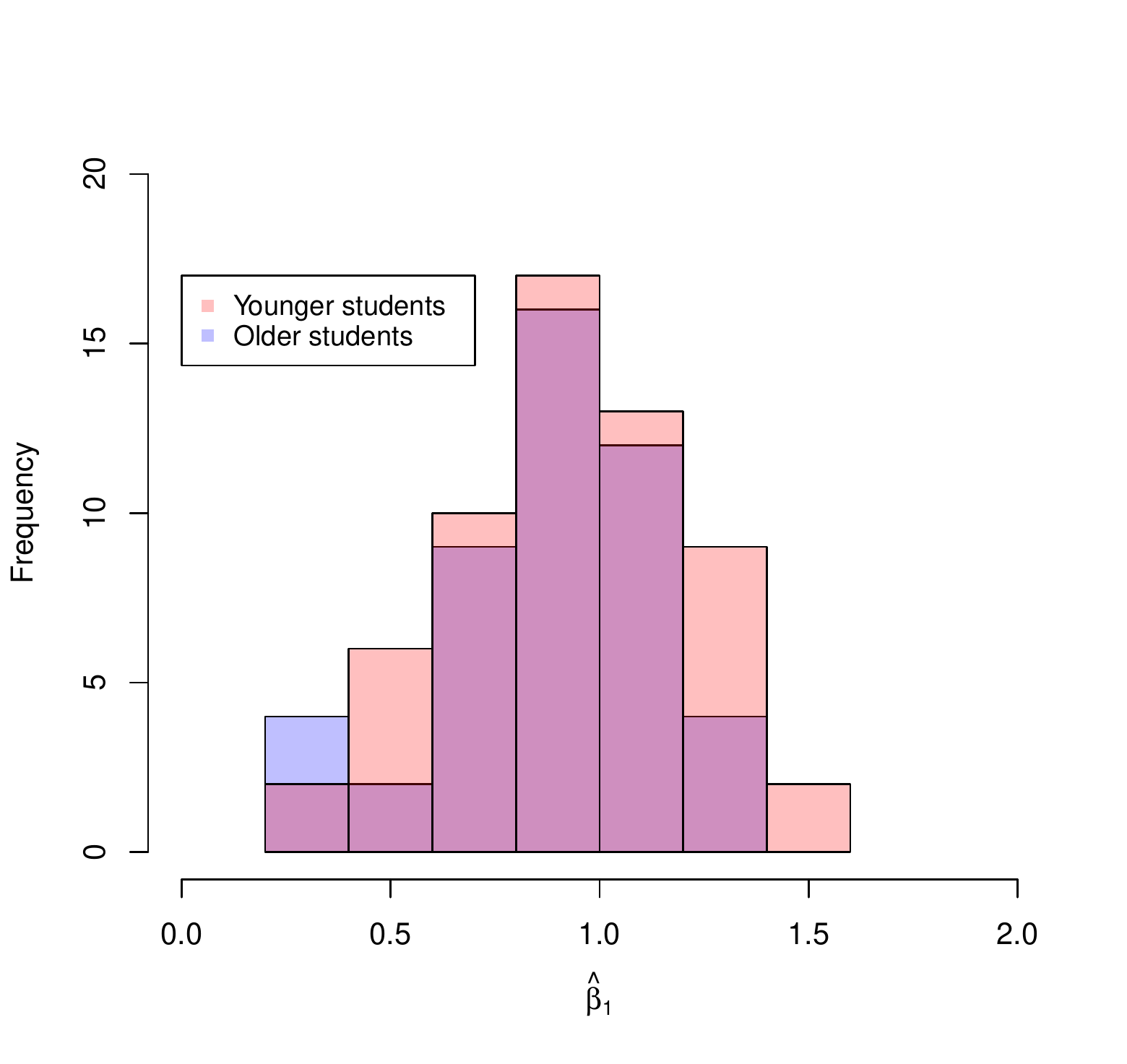}
        \end{minipage}\hfill
        \begin{minipage}[t]{0.48\textwidth}
        \centering
                \includegraphics[width=\textwidth]{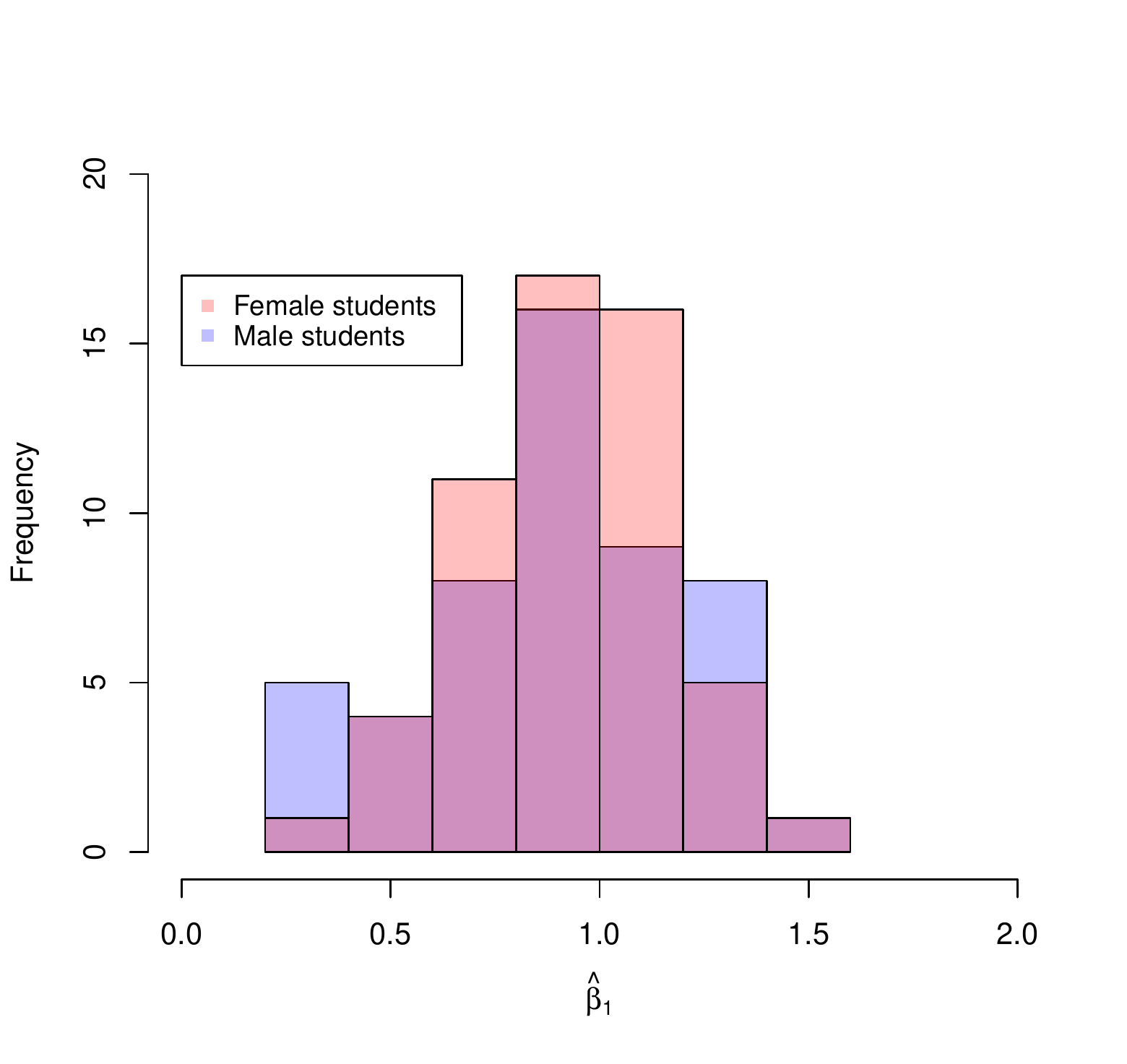}
        \end{minipage}
   \caption{The left-hand figure shows the distribution of $\hat{\beta}_1$, the reaction time and the ability to accelerate and brake, for younger and older students and the right-hand one the distribution for male and female students.}
\label{fig:SteigungYoungOldMaleFemale}
\end{figure}

The slope $\hat{\beta}_1$ of the regression line is related to the stimulus response mechanism connected to reaction time and the ability to accelerate and brake. The comparisons between younger and older and between male and female students can both be considered (see \autoref{fig:SteigungYoungOldMaleFemale}). 
Here, there are virtually no differences. Here, there are virtually no differences. In these cases the mean values and standard errors are $\mu_{old} = 0.87\pm 0.04$, $\mu_{young} = 0.94\pm 0.04$, $\mu_{male} = 0.89\pm 0.04$ and $\mu_{female} = 0.93\pm 0.03$. The previous conclusions can also be confirmed by conducting a t-test. First the normal Q-Q plots showed that the distributions are normally distributed. Then a t-test is used with the null hypotheses, $H_0$ : There is no difference between the groups, i.e. the distributions for younger and older students or for female and male students regarding minimal distances or $\hat{\beta}_1$ are equal. The null hypothesis is only rejected for the comparison between younger and older students at the minimum distance, because p-value $<$ 0.05. Accordingly, only in this case there is no similarity between the groups.

\begin{figure}[htp]
\centering 
        \begin{minipage}[t]{0.4\textwidth}
        \centering
                \includegraphics[width=\textwidth]{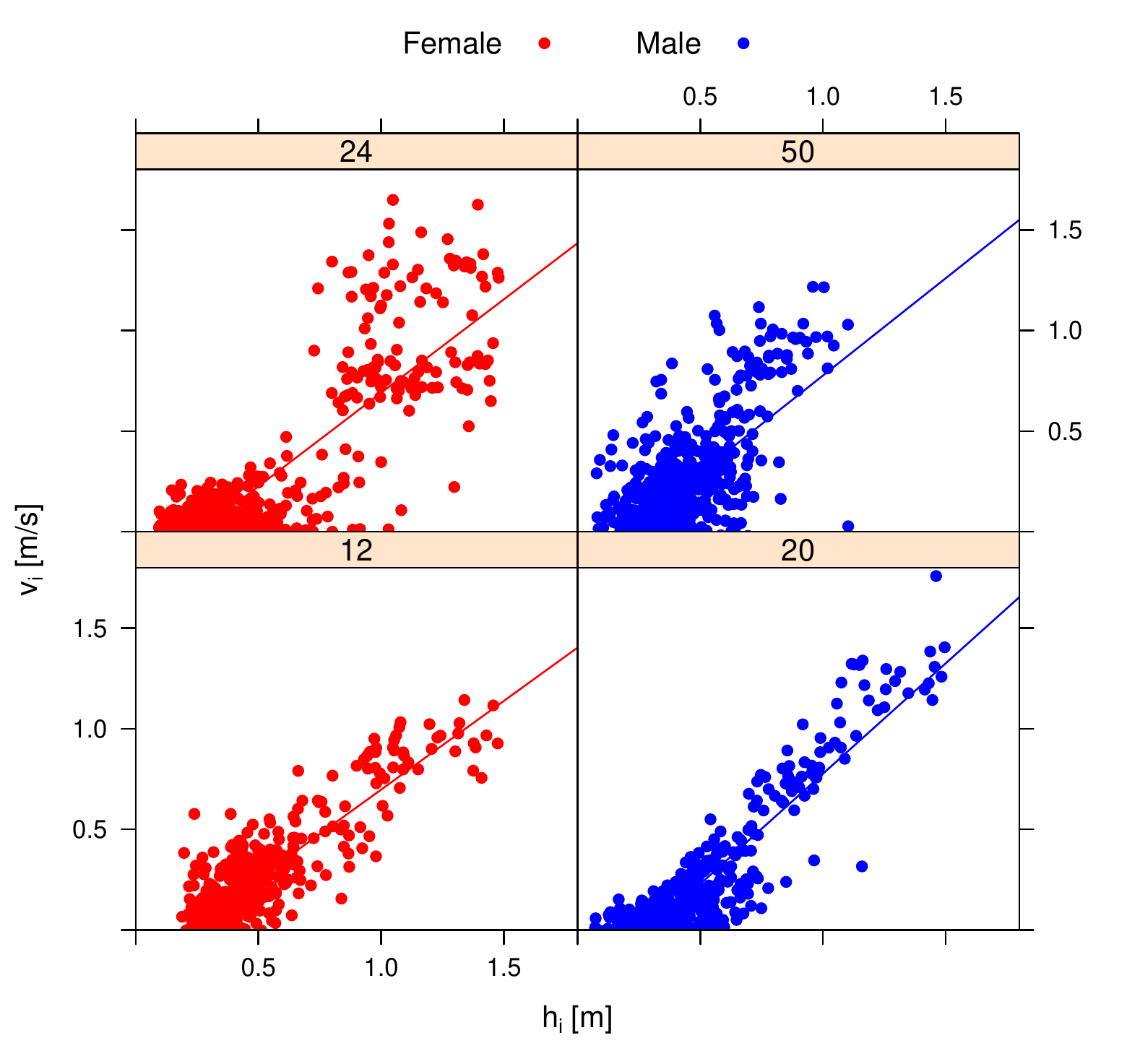}
        \end{minipage}\hfill
        \begin{minipage}[t]{0.4\textwidth}
        \centering
                \includegraphics[width=\textwidth]{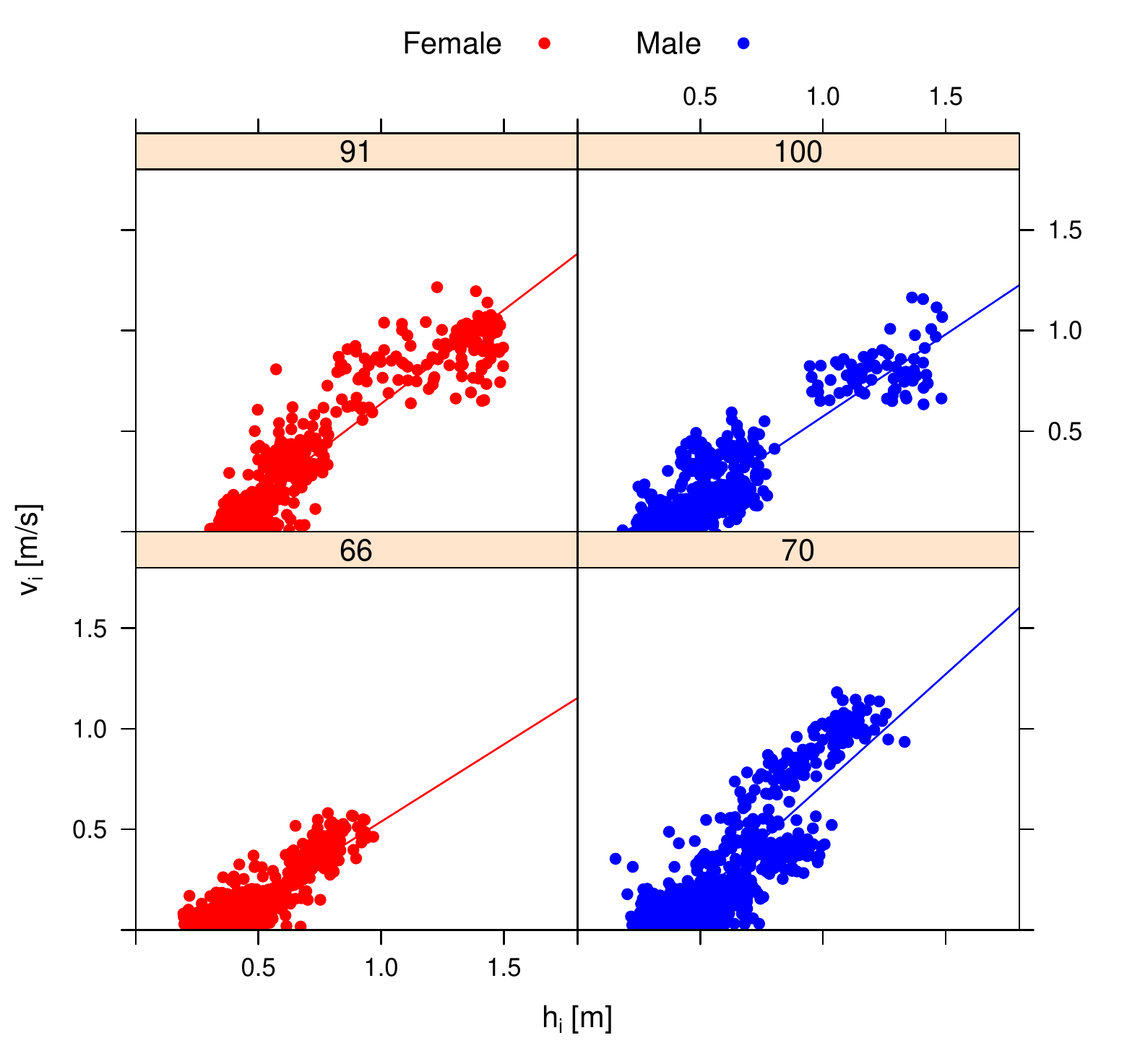}
\end{minipage}  
   \caption{Headway vs. individual speed diagrams with a regression line for representative younger students in grade five on the left-hand side and for older students in grade 11 on the right. These two groups are divided into male and female main IDs, with the numbers in the orange box representing the different main IDs. High and low scattering occurs for each gender and age group.}
\label{fig:HeadwayVelo4IDsWithRegressionBoth}
\end{figure}

Next, the scattering around the regression line is analyzed. When the whole group is divided into younger and older students, as well as into females and males, it becomes clear that in all groups there are headway vs. individual speed diagrams with low and high scattering. In \autoref{fig:HeadwayVelo4IDsWithRegressionBoth}, for each of the two groups, younger and older students, four representative main IDs are selected to illustrate this.

Moreover, the different points represent the measured values $l=1,...,n_i$ for different individuals $i$. The regression line is also shown according to \autoref{eq:SimpleEstimated}. The left-hand figure illustrates the younger pedestrian group and the right-hand one the group of older students. Males are represented in blue and females in red.

\begin{figure}[htp]

\centering 
        \begin{minipage}[t]{0.48\textwidth}
        \centering
                \includegraphics[width=\textwidth]{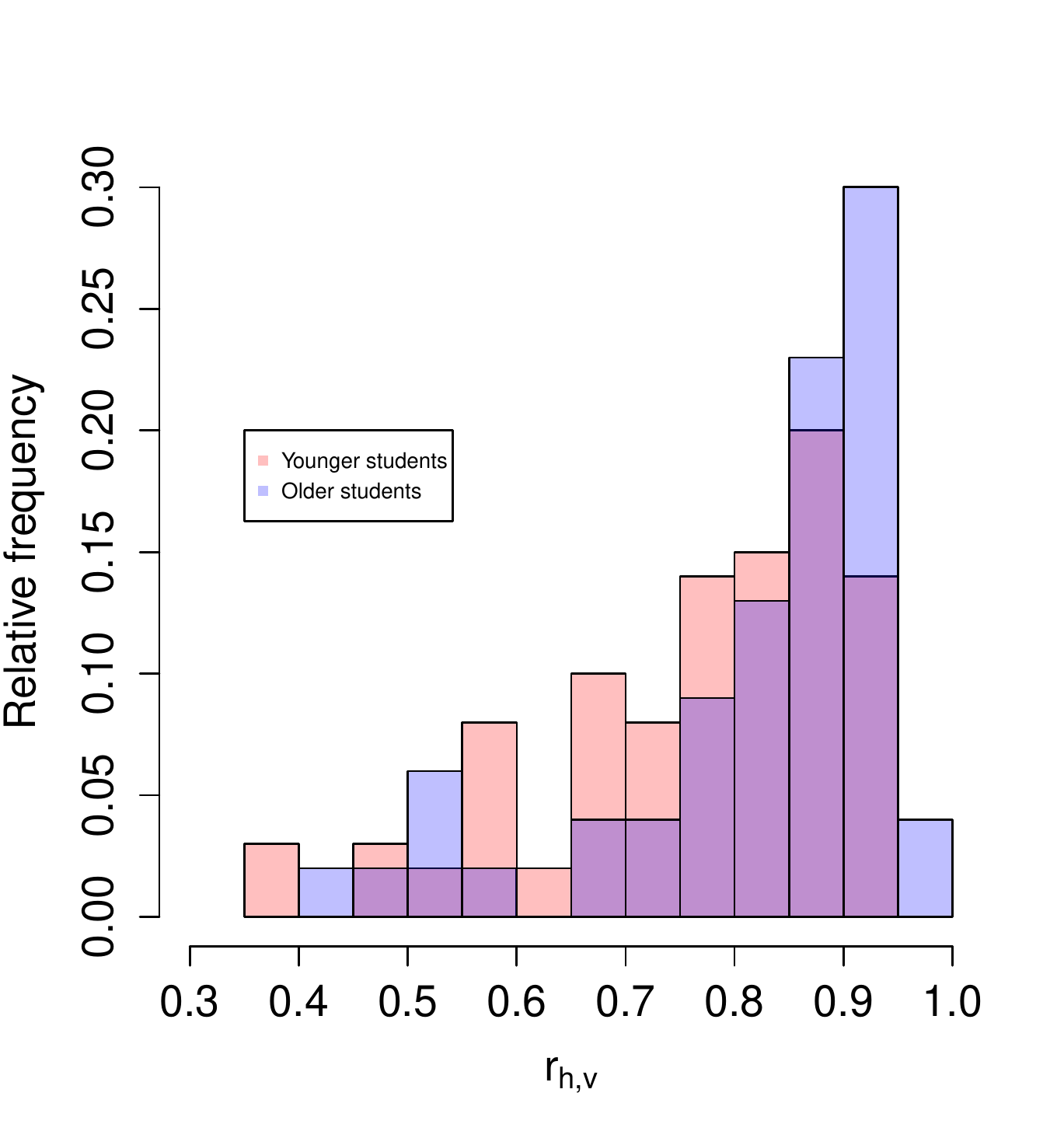}
        \end{minipage}\hfill
        \begin{minipage}[t]{0.48\textwidth}
        \centering
                \includegraphics[width=\textwidth]{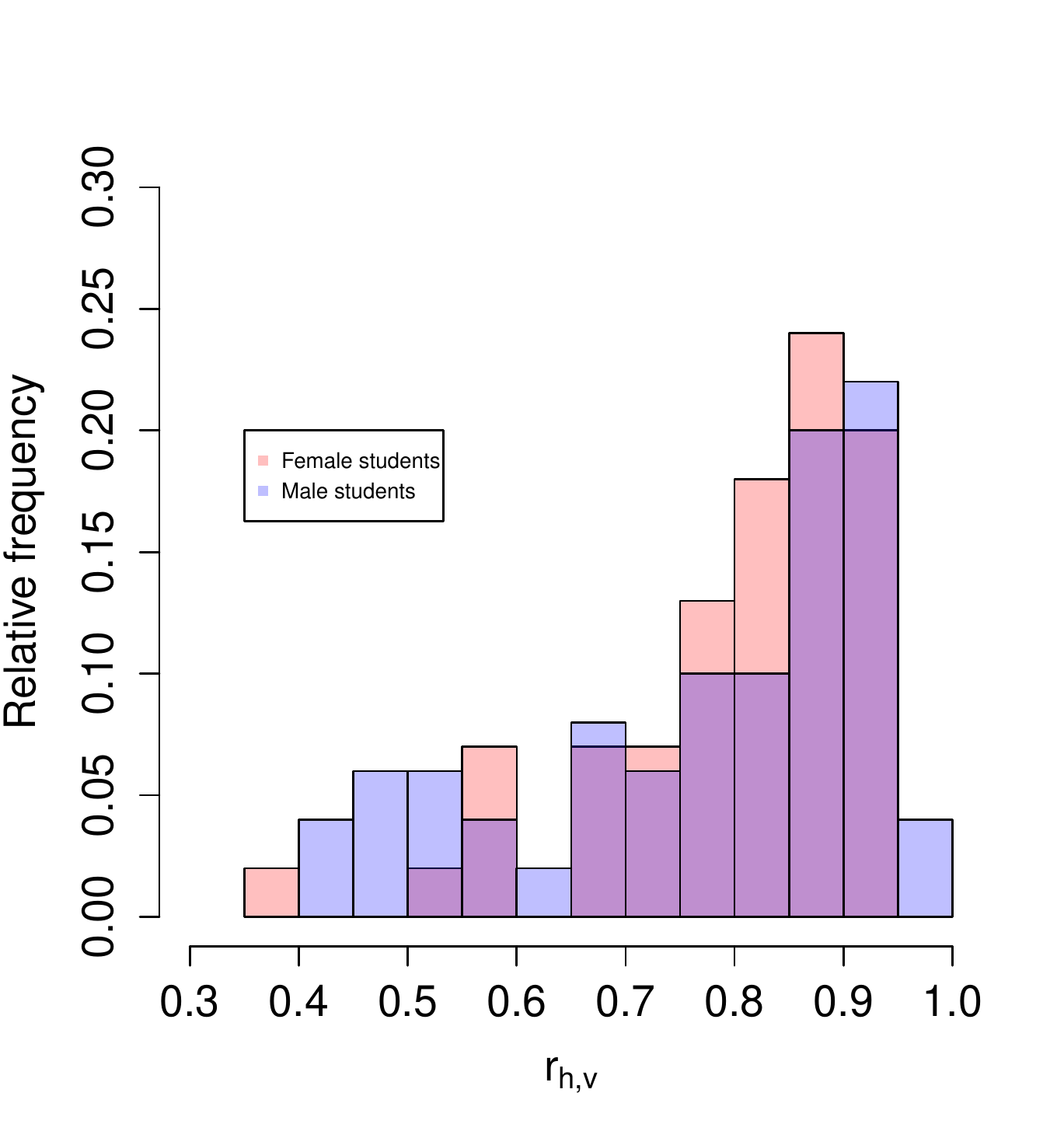}
        \end{minipage}  
     \caption{Distribution of the correlation coefficient between headway vs. individual speed for younger and older students and for female and male students.}
\label{fig:CorrelationGroupsHeadwayVelo}
\end{figure}

In order to examine whether the scattering around the regression line is higher for older or younger students as well as for female or male students, the correlation between the headway and the individual speed is studied (see \autoref{fig:CorrelationGroupsHeadwayVelo}).
\autoref{fig:HeadwayVelo4IDsWithRegressionBoth} shows a tendency for a larger scattering for younger participants and the histograms in \autoref{fig:CorrelationGroupsHeadwayVelo} support this hypothesis.
The mean correlation coefficient \mbox{$\overline{r}_{h,v}$} of the younger group is 0.77 and 0.82 for the older group. When a distinction is made between male and female students, it becomes apparent that for both younger and older students, the group correlation coefficient is so low that there is no significant difference. For the young male students, the value is 0.76 and for the young female students it is 0.77. In contrast to the young group, the values for the older group are 0.80 for the male students and 0.84 for the female students. Considering all female and male students separately, it is apparent that the scattering is slightly larger for male students (see \autoref{fig:CorrelationGroupsHeadwayVelo}). 

In general, the linear regression diagrams representing headway vs. individual speed which illustrate different individuals show that for \mbox{$h < 1.5$ m}, there appears to be virtually no difference between gender and age. It can only be assumed that the scattering is larger among younger students. 
This assumption is confirmed when we look at the histograms (see \autoref{fig:CorrelationGroupsHeadwayVelo}) showing the correlation coefficients of the individuals in the various groups, for instance younger and older students. Furthermore, there is no significant difference between male and female students. A comparison of the minimum distances at $\hat{v}=0$ also shows that there are differences between younger and older students, i.e., distances are greater for older students. When we compare male and female students, the differences are less pronounced. Moreover, a comparison of $\hat{\beta}_1$, the reaction time and ability to accelerate and brake, between younger and older as well as between male and female students shows that there are virtually no differences. 

\subsection{Multiple linear regression}
\subsubsection{Model selection }\label{Model}
In the following, the influence of distance and individual factors such as gender, height and age on speed is investigated. The model structure for the multiple linear regression analysis is explained. The goal is to find a model that takes into account all relevant independent variables for the study of the dependent variable. The individual speed is $v_m$, where $m=1,...,n$ and $n$ is the number of all observations of all individuals. 
At the same time, the model should include as few variables as possible. First, the individual characteristics measured, headway $h$, $age$, $height$, and $gender$ are introduced as independent variables. The variable $alloence$ is used to take into consideration all other unknown individual effects, for example, motivation, attention, or excitement. In addition, the independent variables should not be strongly correlated with each other and should ideally be linked to the speed. 
The variables $height$ and $age$ are strongly correlated ($r_{height,age}=0.89$) and so it is sufficient to include only one of these depending on the research question and the quality of the data, for example granularity. Here, the $height$ is used, since $age$ is only measured as a binary variable with $0$ representing the younger students and $1$ the older ones, while height is categorized according to five levels. It should be noted in this context that there is a correlation between age and height, because the body of the students is still growing, and height can vary between younger and older students. There is either no or only a weak correlation ($r_{x,y}<0.29$) between the other independent variables.
In the following multiple linear models, the variables are considered without units. The first full model that considers all measured individual characteristics is:
\begin{equation}\label{eq:ModelI}
    \mbox{Model \ I: $v_{m}=\beta _0+\beta_1 \cdot h_{m}+\beta_2 \cdot gender_{m}+\beta_3 \cdot height_{m}+\epsilon_{m}$} \ .
\end{equation}
This model allows us to analyze which of the two individual characteristics, height or gender, has a stronger effect on the fundamental diagram. It is important to note that this research question is different from the following question: How strongly do individual characteristics affect the fundamental diagram? To answer this second question, it should be tested whether other variables that would improve Model I and affect the dependent variable $v_{m}$ are ignored.
In comparison to the first model, a further model which takes all other unknown individual effects into account by also including the variable $alloence$ is introduced:
\begin{equation}\label{eq:ModelII}
    \mbox{Model II:  $v_{m}=\beta _0+\beta_1 \cdot h_{m}+\beta_2 \cdot gender_{m}+\beta_3 \cdot height_{m}
    +\sum_{i=1}^{N}\beta_{4i} \cdot alloence_{m}+\epsilon_{m}$} \ , 
\end{equation}
where $\mbox{alloence}_{m}=1$ for all $m$ belonging to individual $i$ and $0$ for all other $m$. $\beta_{4i}$ is an individual coefficient across all measurement points for each student.

In a next step, it is analyzed which of the variables used in \autoref{eq:ModelII} and \autoref{eq:ModelI} should be considered to obtain the best possible model with as few variables as necessary.
One method for making this decision is the model evaluation using Akaike's Information Criterion (AIC) \cite{AIC}. Here, it is decided step by step whether a model improvement can be achieved by omitting independent variables in \autoref{eq:ModelII}. The lower the AIC value, the better the model. This method ends when no further reduction is useful. However, it should be noted that the method does not generally provide absolute criteria for deciding which model is the better choice. By applying it to Model II, the steps of the AIC procedure indicate that $gender$ can be omitted. Interestingly, the same procedure for Model I shows that the model should not be reduced and gender has a significant effect. This evaluation of both models with AIC shows that only taking into consideration of nonmeasured individual characteristics allows a reduction of the model to Model III. Therefore, the variable $alloence$, describing all unknown individual effects, has a significant contribution. This result also shows that Model I does not necessarily include all relevant individual characteristics used to describe the influences on the speed.
\begin{equation}\label{eq:ModelIII}
    \mbox{Model III: $v_{m}=\beta _0+\beta_1 \cdot h_{m}+\beta_2 \cdot height_{m}
    +\sum_{i=1}^{N}\beta_{3i} \cdot alloence_{m}+\epsilon_{m}$}\ .
\end{equation}

\begin{figure}[htp]
\centering
\includegraphics[width=8cm]{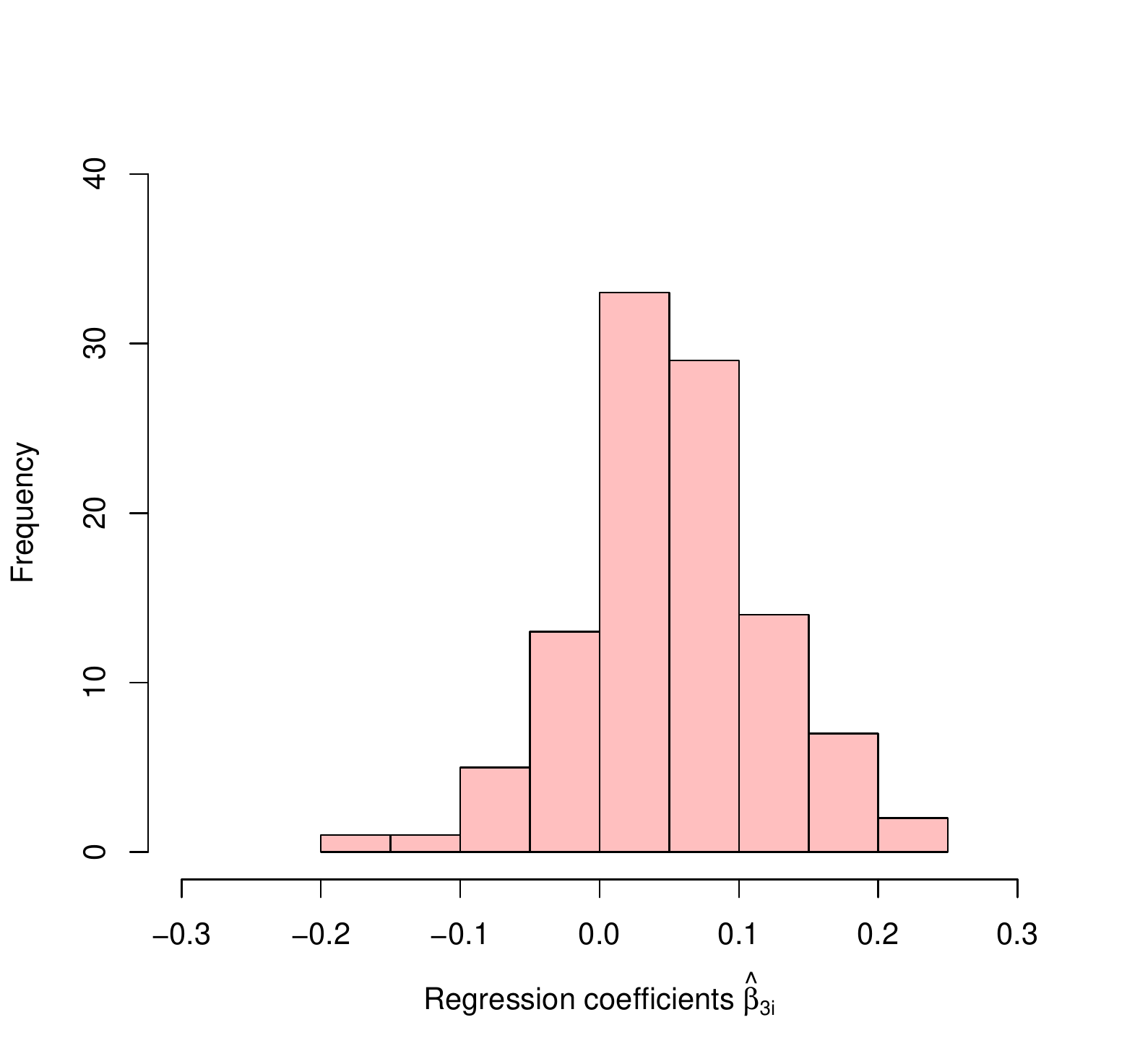}
     \caption{Distribution of the regression coefficients $\hat{\beta}_{3i}$ of all other unknown individual effects for each main ID.}
\label{fig:RegressionCoefficients}
\end{figure}

The result of the Akaike's Information Criterion is consistent with the statement made based on the analysis of the minimum distance, the reaction time, and the correlation between headway and individual speed in section \ref{LinearRegression} above. The difference between younger and older students is greater than between male and female students where there is virtually no difference. Since the difference between men and women is minimum and other individual characteristics predominate, gender can be ignored in Model II. However, Model I does not include the other individual characteristics, so gender cannot be ignored due to its minimum impact. 
Based on Model III, the following estimated regression coefficients are shown in \autoref{eq:ModelEstimated}.
\begin{equation}\label{eq:ModelEstimated}
   \mbox{$\hat{v}=0.23+0.98 \cdot h-0.34 \cdot height
    +\sum_{i=1}^{N}\hat{\beta}_{3i} \cdot alloence$} \ .   
\end{equation}
It can be seen from $\hat{\beta}_0=0.23$, $\hat{\beta}_1=0.98$, and $\hat{\beta}_2=-0.34$ that changes in every predictor variable are significantly associated with changes in speed. The distribution of the values for $\hat{\beta}_{3i}$ are illustrated in \autoref{fig:RegressionCoefficients}. The variable $alloence$ has a positive or negative and stronger or weaker effect on the individual speed in depending on the different regression coefficients $\hat{\beta}_{3i}$ for the individuals. The values are weak but the effect is mostly statistically significant, \mbox{$p \ < \ 0.05$}. In total 83,3\%  are significant. 

In addition, it is also examined whether there are different results or models when the data for younger and older students are analyzed separately. This is not the case. 

\subsubsection{Analysis of Variance (ANOVA)}\label{ANOVA}
\begin{figure}[htp]
    \centering
\includegraphics[width=6.5cm]{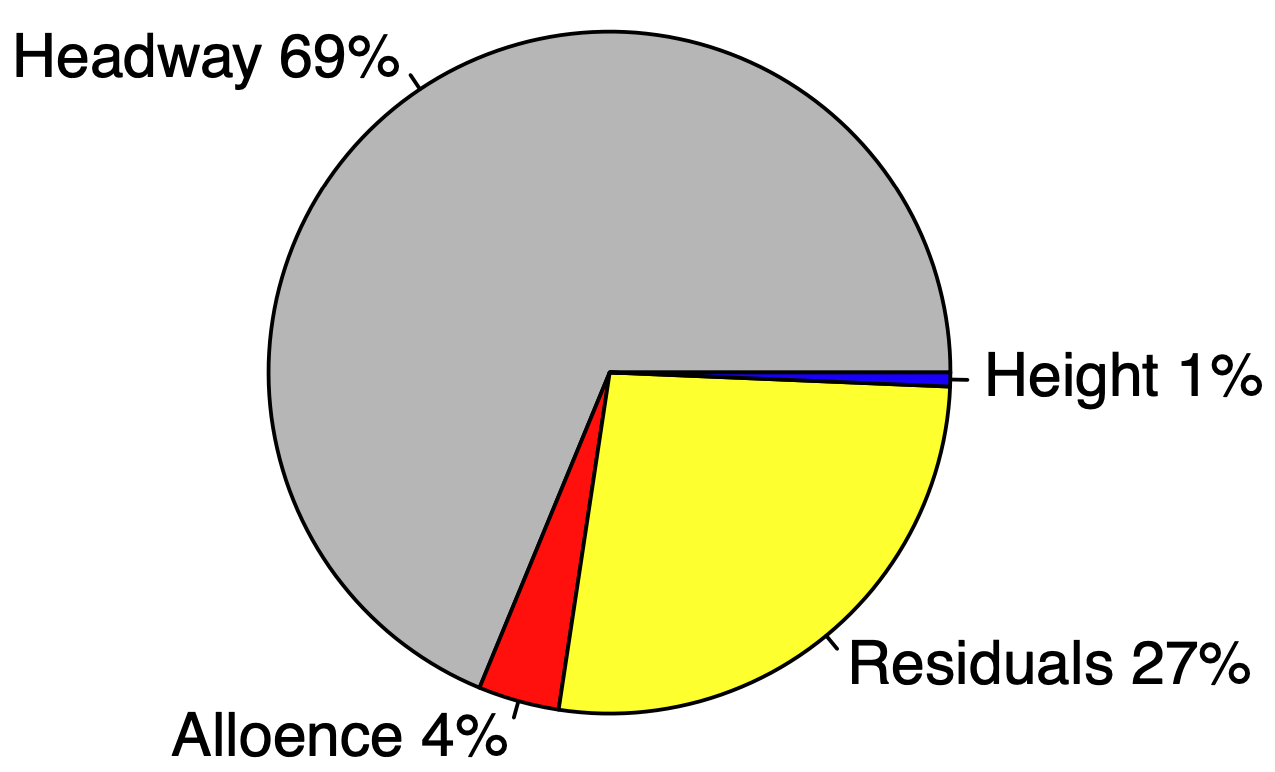}
     \caption{Various effects on the individual speed based on the ANOVA table.}
   \label{fig:AnovaKuchendiagramm}
\end{figure}

Next, Model III (see \autoref{eq:ModelIII}) is used to analyze the effect of the independent variables, headway $h$, $height$, and $alloence$ on the speed. 
On the basis of the ANOVA table, it becomes clear that all variables considered in Model III have an effect on the individual speed. Any effect is significant because the \mbox{$p < 0.05$}.
\autoref{fig:AnovaKuchendiagramm} illustrates the various effects on the individual speed based on the ANOVA table in a pie chart. 
The headway has the largest effect. This is followed by all other unknown individual effects. The height has the lowest effect at $1 \ \%$. The same analysis for Model II (see \autoref{eq:ModelII}) shows that gender has a smaller effect than height. Even without taking headway into account, no larger effect could be attributed to the other independent variables. 

In a further step, the residuals are examined for first-order autocorrelation using the Durbin-Watson test. This test enables us to check whether potentially relevant effects are ignored.
Here, the null hypothesis, $H_0: \ \rho_1=0$, where $\rho_1$ is the theoretical autocorrelation coefficient, implies that the error term is not autocorrelated. If the null hypothesis is rejected, it can also be decided whether positive or negative autocorrelation occurs. If the null hypothesis is rejected, the error term does not fulfill the standard assumptions of the multiple linear regression model. The Durbin-Watson statistics shows a value of $0.7077$. Consequently, the null hypothesis is rejected. In addition, there is a positive autocorrelation due to the low $DW$ value and the limits for the critical values of the Durbin-Watson statistics. Therefore, the headway has the most significant effect on the individual speed, but besides height and all other unknown individual effects there must be other factors that cannot be omitted as an effect on the speed. This indicates that, for example, acceleration phases must be distinguished from braking phases or the stimulus response mechanism connected to reaction time and ability to accelerate, or the locomotion in steps should be modeled more carefully.

\subsection{Mixed Model}
Finally, a mixed model which, unlike the previous models, considers fixed and random factors is used for the analysis. 
Fixed factors are independent variables that are observable, such as headway, height, or gender. The random factors are not observable and may obscure the effect of the factors of interest. These might be individual characteristics included in the variable $alloence$ that are not considered as fixed factors or characteristics such as the attention or motivation of an individual.
This model can be applied to multilevel data, from several observations of an individual or a group and is relevant for the analysis of correlated data. 
Here, the following problem is addressed. The models used previously include the variable $alloence$ as a fixed factor. Now it is used as a random factor. Thus, it is considered as an individual factor which could be correlated with height or gender, too. As shown above, gender has a potentially small effect and is ignored in this analysis.

Two models are compared: a simple model which only includes fixed factors (see \autoref{eq:ModelIII}) and a mixed model. The aim is to establish which model leads to an improvement and should consequently be used. Therefore, a \mbox{$\chi ^2$}-test is performed, which checks whether the mixed model in which the individual effects may obscure the effect of the factors of interest is more efficient than the simple model.
If the null hypothesis is rejected, the mixed model is preferred to the simple model. This is the case if \mbox{$p < 0.05$} applies.
In the present analysis, $H_0$ is rejected so we can conclude that the mixed model in which the individual effects are included as a random factor is preferable to the simple model. Accordingly, it is better to use mixed models for a multivariate analysis in this context: otherwise, the effect of the factors of interest may be obscured. However, one reason could also be the use of the linear model. For some students, the free speed could be included, because this could start earlier than at a headway of 1.5m. 

\section{Conclusions}
The comparison of data provided in the literature for single-file studies considering age as a human factor show that it has an effect on fundamental diagrams. However, a contradictory picture of how age affects the fundamental diagram is found. One possible reason for these contradictions is that the different experiments are not comparable because even if a group is homogeneous in one factor, it could be heterogeneous in other factors such as gender.
To date, only the cumulative data on all individuals in the group have been used for a comparison of the fundamental diagrams. In a new approach, the present study is limited to the simplest system of a one-dimensional single-file school experiment and individual fundamental diagrams are introduced. Moreover, more factors such as height, gender, and age describing the participants on an individual level are measured and used for a multiple regression analysis.  

The focus of the data analysis is on the research question how strongly individual characteristics affect the fundamental diagram. First the research question is about whether individual speed-headway functions show that gender and age have an influence. Using simple linear regression of individual fundamental diagrams, we analyze the minimum distance, the reaction time, and the scattering of the data. The results show that more significant differences occur between older and younger students whereas the differences between male and female students are virtually non-existent. According to this, age has a stronger effect than gender.

Than, the influence of distance and individual factors such as gender, height and age on speed is investigated. For this, model selection steps of the multiple linear regression analysis are performed. The models for the speed depend on $headway$, $gender$, $height$, and $alloence$. The latter is introduced to consider all individual factors which could not be measured, such as motivation or attention. The analysis shows that the variable $alloence$ is crucial and $gender$ could only be omitted in the model when the unknown factor is not ignored.

The analysis of the impact of the variables confirms that the headway is the most crucial factor. This is followed by all other unknown effects and height in particular has a very low percentage. Therefore, the indefinable factors have a greater effect than height or gender. The result is also reflected in the previous statement, as the differences between genders are smaller than between younger and older students. In a further step, an examination of the residuals shows that the individual speed is not only due to the predictor variables used and that potentially relevant effects are ignored. Consequently, besides the height and all other unknown individual effects there must be further effects on the speed that cannot be ignored. When the regression analysis is extended, a simple model that only includes fixed factors and a mixed model that considers the individual speed as a function of the fixed factors headway and height and all other unknown individual effects as a random factor are compared. The analysis shows that the mixed model is preferable. Accordingly, the effect of the individuals should be considered as a random factor. Otherwise, the effect of the factors of interest may be obscured.

For further research, the individual fundamental diagrams can be viewed and analyzed more closely. When we look at the individual fundamental diagrams, it is evident that there are diagrams in which the students keep different distances from different people moving at approximately the same speed.
Future studies could explore whether or not this is because an individual prefers to maintain a certain distances from a certain person. The effect of the people around an individual could then be analyzed on the basis of this information. Moreover, in a further step it can be tried to combine the model e.g. with the model of Tordeux to simulate the single-file pedestrian flow considering individual properties.

\section*{Acknowledgments}
This study is based on a one-dimensional single-file experiment which was performed in 2014 at the school Gymnasium Bayreuther Straße (GBS) in Wuppertal, Germany \cite{Data}. Many thanks to Verena Ziemer who conducted the experiments. The experiments were supported by the German Research Foundation (Grant. No. SE 1789/4-1). We thank the anonymous reviewers whose suggestions and comments helped improve and clarify this manuscript. Funded by the Deutsche Forschungsgemeinschaft (DFG, German Research
Foundation) - 49111148.

\section*{Funding}This research did not receive any specific grant from funding agencies in the public, commercial, or not-for-profit sectors.
\bibliography{ped}
\end{document}